\documentstyle[12pt,aps,epsfig]{revtex}
\newcommand{\be}{\begin{equation}}
\newcommand{\ee}{\end{equation}}
\newcommand{\bea}{\begin{eqnarray}}
\newcommand{\eea}{\end{eqnarray}}
\renewcommand{\d}{{\rm d}}
\newcommand{\PD}{{\partial}}

\def\Journal#1#2#3#4{{#1} {\bf #2}, #3 (#4)}

\def\NP{{Nucl. Phys.}}
\def\PL{{Phys. Lett.}}
\def\PRT{{Phys. Rept.}}
\def\PRL{Phys. Rev. Lett.}
\def\PRD{{Phys. Rev.} D}
\def\PRC{{Phys. Rev.} C}

\def\ZPC{{Z. Phys.} C}

\def\AP{{Ann. Phys. (N.Y.)}}
\def\EPJ{{Eur. Phys. J.}}

\begin{document}
\title
{Calculation of the expansion rate of the three-volume measure in high-energy heavy-ion collisions}
\author{A. Dumitru}
\address{Physics Department, Yale University\\
P.O.\ Box 208124, New Haven, CT 06520, USA\\}
\maketitle   
\begin{abstract}
In ultrarelativistic heavy-ion collisions the local three-volume measure is
expanding in the longitudinal and transverse directions.
This is similar to the Hubble-expansion of the universe, except that the
former is not locally isotropic. As an example the expansion rate is calculated
assuming that the energy-momentum tensor in the central region is that of an
ideal fluid, undergoing Bjorken flow in longitudinal direction, and with
initial conditions as expected for BNL-RHIC energy. While the
longitudinal expansion of three-volume is independent of the energy density of
the fluid, in case of 3+1 dimensional expansion the form of the hydrodynamical
solution (rarefaction wave or deflagration shock) affects the three-volume
expansion rate on the hadronization hypersurface.
As a consequence the average expansion rate on
that surface depends on the transverse size of the system. 
This may reflect in an impact-parameter dependence of the formation
probability of light nuclei and of the freeze-out temperature of
the strong interactions in the system.
\end{abstract}

In high-energy hadron-hadron or nucleus-nucleus collisions secondary
particles are supposedly produced on a so-called proper-time
hyperbola $\tau_0=\sqrt{t^2-z^2}$=const., with their number and energy density
distributions independent of space-time rapidity $\eta=\frac{1}{2}\log
\frac{t+z}{t-z}$~\cite{Bj}. ($z$ and $t$
denote the longitudinal coordinate and the time measured in the global
rest-frame of the reaction, respectively.) 
This should be a reasonable (qualitative) approximation at high energies
and around midrapidity.
In other words, secondary particle production occurs such that invariance
under longitudinal Lorentz boosts, i.e.\ rapidity shifts, is obeyed.
E.g., this holds true at asymptotically high energies for classical non-Abelian
Yang-Mills Bremsstrahlung emitted by sources of color-charge
on the light-cone~\cite{clYM}, if recoil is neglected.
The effect is also built in effective string-models for particle production
at high energies: a hadron emerging from the string-decay with rapidity
$y_p$ gets on mass-shell in the global rest-frame at time
$t=\tau_f\cosh y_p$, where $\tau_f$ is the formation time of the hadron in
its rest-frame~\cite{strAP}. For a discussion within the parton model
see ref.~\cite{PMRD}.

The subsequent dynamical evolution
preserves the invariance of the bulk properties, e.g.\ the energy density
distribution, under longitudinal Lorentz boosts. It is therefore convenient to
switch from $(t,x,y,z)$ to new coordinates, via~\cite{Bj}
\be \label{p_coord}
t = \tau\cosh\eta \quad , \quad
z = \tau\sinh\eta\quad,\quad
x = r_\perp\cos\phi \quad , \quad
y = r_\perp\sin\phi\quad.\nonumber
\ee
The Minkowski line element in terms of the new coordinates is $\d s^2=\d\tau^2-
\tau^2\d\eta^2-\d r_\perp^2-r_\perp^2\d\phi^2$, i.e.\
$g_{\mu\nu}={\rm diag}(1,-\tau^2,-1,-r_\perp^2)$. In the following,
vectors will be written
in the basis were the components 0, 1, 2, 3 correspond to
the $\tau$, $\eta$, $r_\perp$, $\phi$ direction, respectively.
Furthermore, rotational symmetry around the longitudinal direction is assumed.

Given some boundary conditions on the hypersurface
$\tau=\tau_0$, the evolution within the forward
light-cone is governed by~\cite{MTW_G}
\be \label{ContEqs}
\PD\cdot T=0 \quad,\quad
\PD\cdot N_B = 0\quad.
\ee
In  the ideal fluid approximation (natural units are employed,
$\hbar=c=k_B=1$)
\be \label{emTensor}
T^{\mu\nu} = (\epsilon+p)u^\mu u^\nu -pg^{\mu\nu}\quad,\quad
N_B^\mu = n_B u^\mu \quad.
\ee 
The four-velocity of the locally comoving
frame is normalized to $u\cdot u=1$, and is given by
$u^\mu=\gamma_\perp (1,0,v_\perp,0)$,
where $\gamma_\perp^{-2}=1-v_\perp^2$~\cite{Baym,trexp,RiGy,DumRi}.
The projections of the equations for the energy-momentum tensor
parallel and orthogonal to $u$, and the continuity
equation for the net baryon current yield
\bea
u\cdot\PD\, \epsilon &=&
       - \left( \epsilon+p\right) \PD\cdot u\quad,\label{epseq}\\
\PD_\eta p &=& 0\quad,\label{peq}\\
u\cdot\PD\, n_B &=&
       -  n_B\, \PD\cdot u\quad,\label{neq}
\eea
with
\be \label{udotd}
u\cdot\PD = \gamma_\perp\left( \PD_\tau+v_\perp\PD_\perp\right)\quad.
\ee
The expansion scalar $\PD\cdot u$ is given by
\be \label{du}
\PD\cdot u \equiv
\frac{1}{\sqrt{-g}}\PD_\mu \left(\sqrt{-g} u^\mu\right) =
\frac{\PD \gamma_\perp}{\PD\tau} +
\frac{\gamma_\perp}{\tau} + \frac{u_\perp}{r_\perp} +
\frac{\PD u_\perp}{\PD r_\perp}\quad.
\ee
Thus, for purely longitudinal expansion ($u_\perp=0$, $\gamma_\perp=1$), we
simply have $\PD\cdot u=1/\tau$. This is obvious because
in this case $V=\pi R_\perp^2 \tau$ is the three-volume on a
$\tau={\rm const.}$ 
hypersurface corresponding to a length in space-time rapidity of
$\Delta\eta=1$. In other words,
the Hubble constant in longitudinal direction is
\be
H(\tau)=\frac{1}{\tau}\quad,
\ee
for $\tau\ge\tau_0$ and $r_\perp\le R_\perp$. $H(\tau_0\sim0.5-1$~fm$)$ is
roughly $10^{40}$ times bigger than the present rate of expansion of the
universe, and about $10^{18}$ times the Hubble constant at the cosmological 
hadronization phase transition.
Note that for purely longitudinal expansion $\PD\cdot u$ (or $H$) is
independent of $\epsilon$ and $n_B$, i.e.\ the fluid does not influence the
expansion rate. This is due to eq.~(\ref{peq}): free (non-accelerated) flow
always means $\eta={\rm const.}$, {\em independently} of the equation
of state $p(\epsilon,n_B)$.

As a side-remark, note that for {\em spherically} symmetric,
three-dimensional boost-invariant expansion, $\vec v=\vec r/t$.
One finds $\PD\cdot u=3/\tau$, where~(\ref{p_coord}) is
now replaced by $t = \tau\cosh\eta$,
$r = \tau\sinh\eta$, that is $\tau^2=t^2-r^2$ and
$\eta=\frac{1}{2}\log\frac{t+r}{t-r}$.
Again, the expansion scalar $\PD\cdot u$ is independent of the
equation of state.

One also observes from eqs.~(\ref{epseq},\ref{neq}) that the continuity
equations (\ref{ContEqs}) can not be extrapolated to $\tau=0$ because
the three-volume vanishes (the Jacobian
$|(\PD(t,z)/\PD(\tau,\eta)|=\tau$ of~(\ref{p_coord}) is zero).
Nevertheless, the transformation from
$(t,z)$ to $(\tau,\eta)$ is not useless, because the classical description
breaks down for $\tau\rightarrow0$ anyway.
In such reactions the $\langle p_\perp\rangle$ of
produced quarks and gluons are on the order of 1~GeV, and thus the uncertainty
relation sets a time-scale of $\sim1/\langle p_\perp\rangle=0.2$~fm, below
which this approach must fail.

If transverse expansion is superimposed on longitudinally boost-invariant
expansion, $\PD\cdot u$ of course
depends on the transverse flow velocity, cf.\ eq.~(\ref{du}), which
by itself depends on transverse pressure gradients (caused by energy density
and/or baryon density gradients). Thus, the
evolution of $\epsilon$, $n_B$, $u$, and $\PD\cdot u$ is coupled via
eqs.~(\ref{epseq},\ref{neq}).
Consider, in particular, a hypersurface $\sigma^\mu=(\tau,\eta,r_\perp,\phi)$
in space-time (e.g.\ a surface of constant time, or temperature, or the
hadronization hypersurface
$\lambda=0$, where $\lambda$ denotes the local fraction of quark-gluon phase).
In parametric representation, $\sigma^\mu$
is a function of three parameters~\cite{MTW_G}.
In our case, due to the symmetry under rotations around and Lorentz-boosts
along the beam axis, two of these parameters can simply be identified with
$\eta$ and $\phi$, while $\tau$ and $r_\perp$ depend only on the third
parameter, call it $\zeta$. Thus, $\zeta\in\left[0,1\right]$ parametrizes the
hypersurface in the planes of fixed $\eta$ and $\phi$ (counter clock-wise).
Then, the normal is\footnote{In our basis
$\epsilon_{\mu\alpha\beta\gamma}$ contains a factor
$\sqrt{-g}=r_\perp\tau$.}
\be
\d\sigma_\mu = \epsilon_{\mu\alpha\beta\gamma}
\frac{\PD\sigma^\alpha}{\PD \zeta}
\frac{\PD\sigma^\beta}{\PD \eta}
\frac{\PD\sigma^\gamma}{\PD \phi}
\d\zeta\d\eta\d\phi
= r_\perp\tau\left(-\frac{\PD r_\perp}{\PD \zeta},0,
\frac{\PD\tau}{\PD\zeta},0\right) \d\zeta\d\eta\d\phi
\ee
The three-volume measure on the hypersurface is
\be \label{hs3dV}
\d V\equiv\d\sigma\cdot u = r_\perp \tau \left( u_\perp\frac{\PD\tau}{\PD\zeta}
-\gamma_\perp\frac{\PD r_\perp}{\PD \zeta}\right)\d\zeta\d\eta\d\phi\quad.
\ee
E.g., on constant-$\tau$ hypersurfaces and for purely longitudinal flow,
$\d V=\tau\d^2r_\perp\d\eta$.
Note that $\PD\cdot u$ is simply the rate of expansion of
the three-volume measure~(\ref{hs3dV}), since
\be \label{Vexp}
u\cdot\PD\, \d V = \d V \, \PD\cdot u
\quad.
\ee
Eq.~(\ref{Vexp}) can be verified by an explicit calculation using
the relations~(\ref{udotd},\ref{du},\ref{hs3dV}). However, the following proof
is simpler and more general (it does not assume longitudinal boost-invariance
and cylindrical symmetry). Note that the total net baryon number
\be
{\cal B}\equiv\int \d\sigma_\mu N_B^\mu =
\int \d V n_B
\ee
is a constant, and thus $u\cdot\PD{\cal B}=0$. Therefore,
\be
\int \left(u\cdot\PD\d V\right) n_B =
u\cdot\PD \int \d V n_B - \int \d V u\cdot\PD n_B
= \int \d V n_B \left(\PD\cdot u\right)\quad.
\ee
In the second step, the continuity equation~(\ref{neq}) has been used.
This must be true for any arbitrary function $n_B$, and therefore the
identity~(\ref{Vexp}) must hold.

To discuss a specific example, equations~(\ref{ContEqs}) have been
solved numerically employing the finite-difference scheme RHLLE, as described
and tested in~\cite{RHLLE}. The initial conditions
were chosen as might be appropriate for collisions of heavy
nuclei at BNL-RHIC energy, $\sqrt{s}=200$~GeV per incident nucleon
pair. The main goal~\cite{HaMu} of these experiments is to repeat, in the
laboratory, the QCD phase transition that occured at some stage in the
universe. As already mentioned above, the difference is that the
``Hubble-constant'' is much larger in high-energy heavy-ion collisions.
The energy density is expected~\cite{HaMu,therm} to be significantly higher
than in lower-energy
reactions, such that the expansion effect, and the influence of the QCD
hadronization phase transition, should be more prominent.

In particular, on the $\tau=\tau_0$ hypersurface, an entropy per net
baryon of $s/n_B=200$ is assumed; see~\cite{DumRi}
for the resulting single-particle transverse momentum spectra
as well as average transverse momenta and velocities of various hadrons.
For simplicity, $s$ and $n_B$ are assumed to be homogeneously distributed
on the $\tau=\tau_0$ hypersurface, independent of $\eta$ and $\phi$,
and proportional to a step function, $\Theta\left( R_\perp-r_\perp\right)$.
Except close to the light-cone, where the flow velocities are largest but where
the fluid is very dilute, the bulk dynamics is not very sensitive to
the precise initial profile~\cite{DumRi}.

Estimates for $\tau_0$ at RHIC energy span the range $0.2-1$~fm~\cite{therm}.
Here, the average value of $\tau_0=0.6$~fm is employed.
For the given initial conditions, and for the MIT bag-model
equation of state (EoS) described below, the initial temperature is
$T(\tau_0)\approx300$~MeV.

To close the system of equations (\ref{ContEqs}) one has to specify
an EoS, which determines the function $p(\epsilon,n_B)$. In the low-temperature
region a gas of non-interacting hadrons that includes all
known~\cite{PDt} strange\footnote{The strangeness chemical potential is
determined by the requirement that the net strangeness density $n_S$, and thus
the strangeness current $N^\mu_S\equiv n_S u^\mu$, vanish
everywhere.} and non-strange hadrons and hadronic resonances up
to a mass of 2~GeV is assumed. At high temperatures, the EoS is that of an
ideal
gas of quarks, antiquarks (with masses $m_u=m_d=0$, $m_s=150$~MeV), and gluons.
In addition, a bag term~\cite{MITbag} $+B$ in the energy density,
and $-B$ in the pressure is included. This is the same as adding a term
$Bg^{\mu\nu}$
to the energy-momentum tensor of the quark-gluon fluid. Mathematically,
the bag term resembles the cosmological-constant term which is
introduced in some cosmological models~\cite{MTW_G}. However, since it
is added to the energy-momentum tensor of the quark-gluon fluid only, and
not to that of the hadronic fluid, it changes the form of the hydrodynamical
solution (deflagrations can occur, see below).

Finally, the phase coexistance region is obtained from Gibbs' conditions of
phase equilibrium. Thus, the EoS exhibits a first-order phase transition.
The value of
$B$ is determined by $T_C$; for $T_C=160$~MeV, which is roughly in accord with
recent lattice-QCD results~\cite{LQCD}, one finds $B=380$~MeV/fm$^3$.

\begin{figure}[htp]
\centerline{\hbox{\epsfig{figure=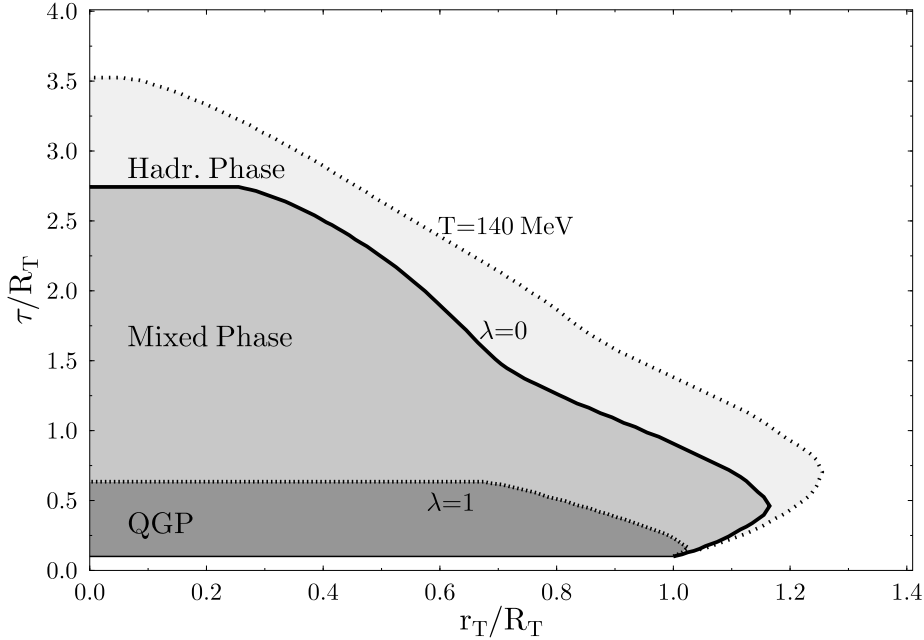,height=9cm}}}
\caption{Hypersurfaces corresponding to $\lambda=1$ (boundary
between QGP and mixed phase),
$\lambda=0$ (boundary between mixed phase and pure hadron phase,
i.e.\ the hadronization hypersurface), and the $T=140$~MeV isotherm;
for $\tau_0/R_\perp=1/10$.}
\label{hyperfig}
\end{figure}  
Fig.~\ref{hyperfig} shows the hypersurfaces where the
pure QGP and the mixed phase end, respectively, as well as the $T=140$~MeV 
isotherm. 
One observes a rarefaction wave developping at $r_\perp=R_\perp$,
$\tau=\tau_0$. It accelerates a part of the quark-gluon fluid
in transverse direction, before
the boundary between quark-gluon plasma and mixed phase is reached
($\lambda=1$). Those parts of the system which have not been affected
by transverse expansion enter the mixed phase at the same time, which
is the horizontal part of the $\lambda=1$ hypersurface at $\tau/R_\perp
\simeq0.6$. The expansion rate within those space-time regions is simply
$1/\tau$, as discussed above.

During the transition to hadronic matter, entropy
is converted from the quark-gluon to the hadronic phase at a rate which
of course depends on $\PD\cdot u$.
For example, assume for simplicity a net baryon free fluid, and that
shock-solutions do not occur,
such that entropy is conserved. Then, the continuity equation for the entropy
four-flow, $\PD\cdot(su)=0$, yields
\be \label{s_conv}
u\cdot\PD\, \lambda = - \left(\PD\cdot u \right) \left( \lambda +
\frac{s^H}{s^Q-s^H} \right)\quad.
\ee
$s^H$, $s^Q$ denote the entropy densities of the hadronic and quark-gluon
fluids at $T=T_C$. $\lambda$ is the local fraction of quarks and gluons within
the mixed phase, such that
the total entropy density is $s=\lambda s^Q + (1-\lambda)s^H$.
Eq.~(\ref{s_conv}) shows that the rate of adiabatic conversion of quark-gluon
fluid into hadronic fluid is governed by the three-volume expansion rate
$\PD\cdot u$, which is larger than $1/\tau$ in the space-time regions
where transverse expansion is active. Therefore, hadronization is faster
close to the surface, $r_\perp\sim R_\perp$, than in the interior,
cf.\ Fig.~\ref{hyperfig}. In particular, the local expansion rate of the
hadronization volume ($\lambda=0$ hypersurface) determines whether the
emerging hadrons can maintain local equilibrium or not.

\begin{figure}[htp]
\centerline{\hbox{\epsfig{figure=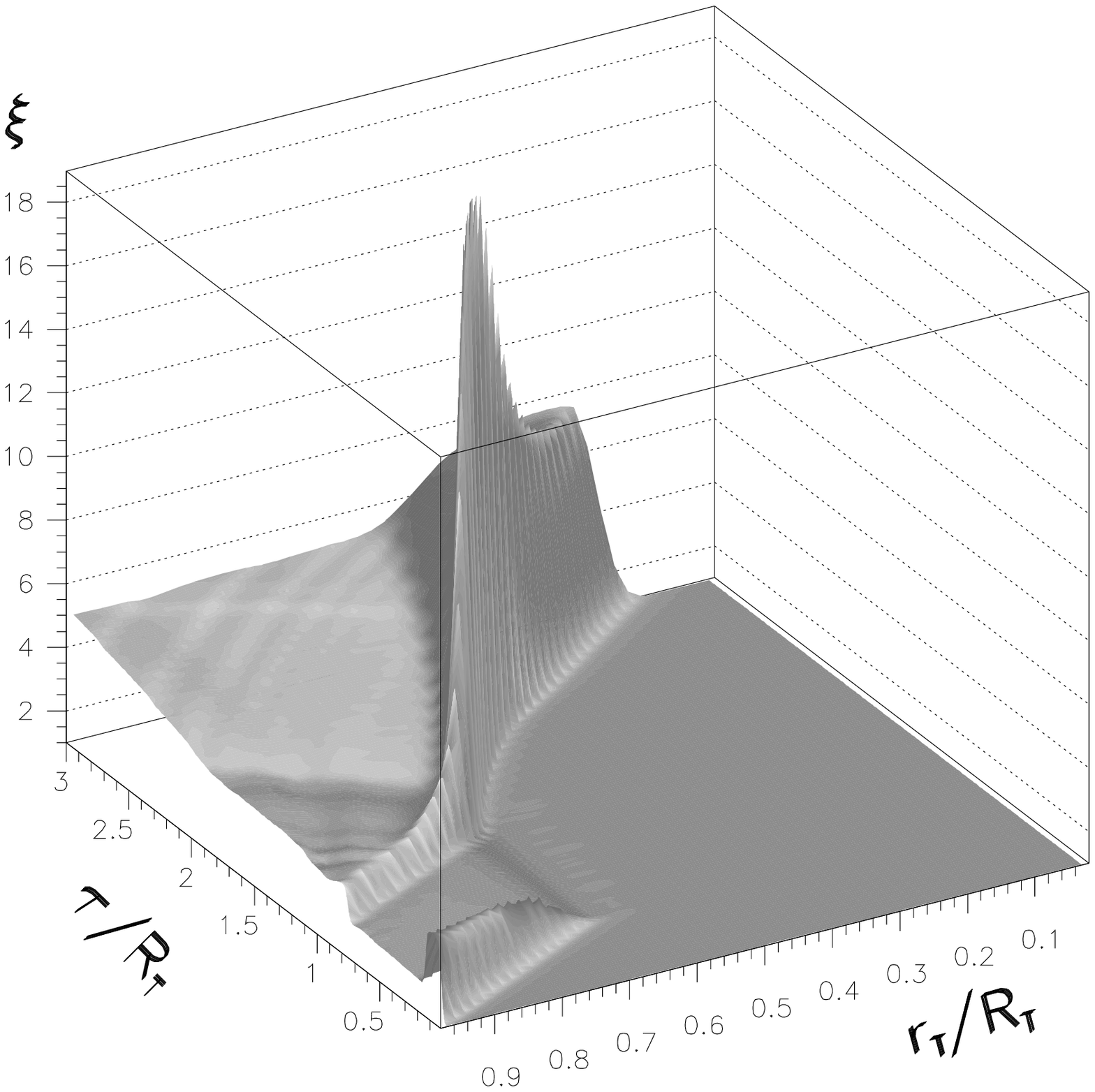,height=9cm}\epsfig
{figure=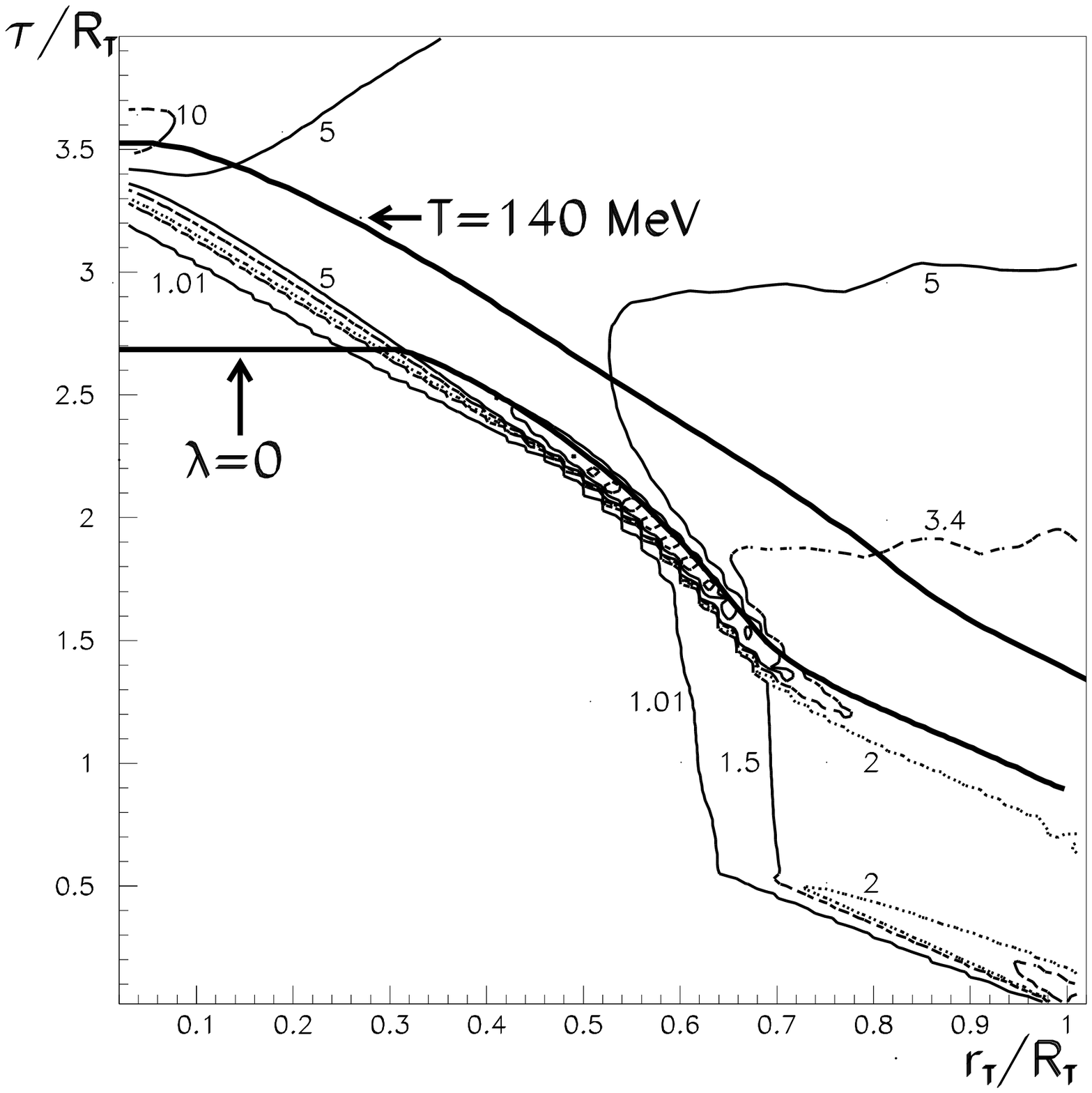,height=9cm}}}
\caption{Left: the expansion factor $\xi\equiv\tau\PD\cdot u$ as a function
of $\tau$ and $r_\perp$. Right: curves of constant $\xi$,
labeled by the value of $\xi$ to which they correspond to; the thick lines
depict the hadronization hypersurface, $\lambda=0$, and the $T=140$~MeV isotherm,
respectively. Both for $\tau_0/R_\perp=1/10$.}
\label{dufig_r6}
\end{figure}

Fig.~\ref{dufig_r6} shows the expansion factor $\xi\equiv
\tau\PD\cdot u$ as a function of $\tau$ and $r_\perp$.
$\PD\cdot u$ is multiplied by $\tau$ to have $\xi=1$
in space-time regions where the expansion is purely longitudinal.
In the front-left corner one can see the expansion induced by the
rarefaction in the quark-gluon fluid. Once the mixed phase is reached,
the rarefaction wave ''stalls''.

This is due to the very small velocity of sound in the phase coexistance region.
For recent discussions
of the consequences of this effect in cosmology see e.g.~\cite{vs_cosm}.
In heavy-ion collisions, it leads to ``stall'' of the transverse
expansion~\cite{trexp,RiGy,DumRi}. Fig.~\ref{dufig_r6} shows that therefore
$\xi$ is nearly time independent until hadronization is completed and the
purely hadronic phase is reached ($\lambda=0$): the curves of constant
expansion rate (right panel of Fig.~\ref{dufig_r6}) at $r_T\approx0.6$ and $0.7$
are nearly parallel to the time-axis.

The hadronization hypersurface ($\lambda=-0$, i.e.\ infinitesimally smaller than 0)
and the $T=140~$MeV isotherm
from Fig.~\ref{hyperfig} are shown again in Fig.~\ref{dufig_r6}. Clearly,
the expansion rate on these hypersurfaces is {\em not constant}, i.e.,
they are not hypersurfaces of homogeneity. This is in contrast
to an isotropically expanding homogeneous universe~\cite{DJS}. Thus, in
very high-energy heavy-ion collisions it does not seem natural to assume
decoupling on a hypersurface of constant temperature, as already
pointed out in~\cite{hung98a}.

For the equation of state with a first-order phase transition, hadronization is
accomplished via a deflagration shock~\cite{trexp,RiGy,RHLLE,defl}.
(For the geometry at hand, the shock front is
actually curved, as can be seen in Fig.~\ref{dufig_r6}).
The final-state of the
shock is the steady-state corresponding to the Chapman-Jouget
point~\cite{RHLLE,ZRdsh} (maximum entropy production). The hadronization point
$\lambda=0$, however, is in general located on the shock
(and has higher pressure than at the CJ-point).
On the shock front $\xi\rightarrow\infty$ for an ideal fluid. In reality,
a finite viscosity will smear out the shock front slightly; in the figure,
$\xi$ remains finite because in the numerical solution the derivatives in
eq.~(\ref{du}) are replaced by finite differences.
Behind the shock front the expansion proceeds again via simple rarefaction waves,
and $\xi$ is finite.

The large expansion rate of the hadronization three-volume could open
the interesting possibility that in the vicinity of that hypersurface
the fluid breaks up into smaller droplets~\cite{Mis99} which decouple
from each other. This
can lead to rapidity fluctuations~\cite{Mis99,HeiJ} or even the
formation of droplets of disoriented chiral condensate~\cite{Scav}.
However, this scenario will not be discussed here in detail.

\begin{figure}[htp]
\centerline{\hbox{\epsfig
{figure=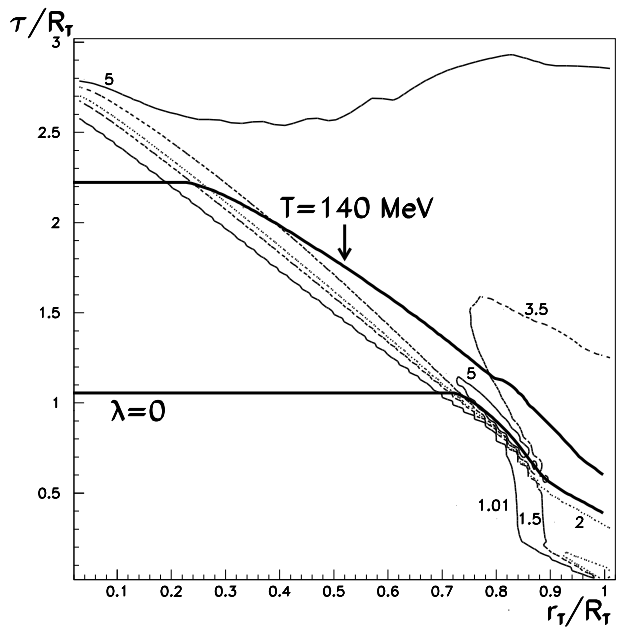,height=9cm}}}
\caption{Curves of constant $\xi$ (for $\tau_0/R_\perp=1/25$, i.e.\ larger
system than in Fig.~\protect\ref{dufig_r6})
labeled by the value of $\xi$ to which they correspond to; the thick lines
depict the hadronization hypersurface, $\lambda=0$, and the $T=140$~MeV isotherm,
respectively.}
\label{dufig_r15}
\end{figure}  
Clearly, those three-volume elements of the hadronization hypersurface that
lie on the shock front will expand rather rapidly, even if the shock-width
were slightly smeared out by a finite viscosity.
It is interesting to note that for the smaller initial radius one obtains
$\xi>1$ almost over the entire hadronization hypersurface (and behind it),
while for large $R_\perp$, cf.\ Fig.~\ref{dufig_r15},
$\xi=1$ over a large part of the hadronization hypersurface.
For very large systems the expansion scalar $\PD\cdot u$ approaches the
''trivial'', EoS independent form $\PD\cdot u=1/\tau$.

Of course, the question arises how the large expansion rate
affects the observed spectra of hadrons.
The most pronounced effect would be expected in space-time regions where
the expansion scalar is large and the fluid is already rather dilute.
E.g., the decoupling from local thermal equilibrium should be
influenced~\cite{hung98a,sorge2,locexp}: a large $\PD\cdot u$ hinders
scattering between the particles of the fluid.
Therefore, the produced hadrons should decouple at
higher $T$ as compared to the case of small expansion rate,
cf.\ also the discussion in~\cite{hung98a,sorge2}.
In other words, the strong interaction freeze-out hypersurface should approach
the hadronization hypersurface as beam energy increases
(as observed in~\cite{DuBass}) and as the initial radius decreases.

If indeed the strong interactions cease at higher temperature as compared to
a slowly expanding three-volume measure,
the average phase-space density of frozen-out pions
or kaons should also increase. This quantity can be estimated from
measurements of the two-particle correlation function~\cite{pi_psd}.
Moreover, the synthesis of light nuclei (and anti-nuclei) might
be suppressed if the three-volume measure expands rapidly,
because it becomes more difficult to coalesce (anti-)nucleons into clusters.
Also, unlike in a static fireball in global thermodynamical equilibrium
one may expect
that the decoupling conditions of various hadron species differ:
a large volume expansion rate should lead to earlier decoupling of hadron
species with small cross section~\cite{DuBass}.
Finally, if the QCD chiral phase transition is of first-order only
at larger baryon-chemical potential $\mu_B$~\cite{tricrit},
but a smooth cross-over at small
$\mu_B$, the hydrodynamical solution (and thus the
expansion rate) might look very different on the two sides of the
critical point. In the region where the phase transition is
first order, a shock can develop and lead to a rather large expansion rate.
In the cross-over region, on the other hand, only rarefaction waves but no
shocks will form, and the average expansion rate at
chiral symmetry restoration should be
smaller. As discussed above, this difference might reflect in the freeze-out
properties of the strongly interacting system.

To analyze these points in detail, one has to supplement the fluid-dynamical
solution on the hadronization hypersurface with a more detailed kinetic
treatment of the hadronic stage, which explicitly
accounts for the various (elastic
and inelastic) elementary hadron-hadron scattering processes~\cite{DuBass}.
One might then be able to study the effect of the three-volume expansion
on the evolution and freeze-out properties of the hadron fluid. 
Those calculations found that some hadrons are indeed
emitted directly from the hadronization hypersurface without scattering
any further, and that the large expansion rate almost ''freezes'' the
chemical composition of the hadron fluid.
More detailed studies, e.g.\ with varying $\tau_0/R_\perp$,
will be performed in the future.

Similar studies could be done in the laboratory by varying the initial
transverse size.
One should keep in mind, however, that the energy density in the central
region decreases with mass number, which could partly counterbalance
the effect.

In summary, transverse expansion couples the scalar $\PD\cdot u$, where $u$ is
the four-velocity of the locally comoving frame, and which can be interpreted
as the expansion rate of the local three-volume measure,
to the properties of the fluid (energy and baryon density and the EoS).
If hadronization proceeds via a shock wave, the expansion rate
can become particularly large on the hadronization hypersurface. The hadrons
produced on such a shock may decouple immediately.
The average expansion rate on the hadronization hypersurface
depends on the size of the system (at fixed initial energy density).
This may reflect in a system-size dependence of the coalescence
probability of light (anti-)nuclei and of the freeze-out properties of
the strongly interacting system (average phase space density and temperature).

{\bf Acknowledgements:}
I thank the Yale Relativistic Heavy Ion Group for kind hospitality and
support from grant no.\ DE-FG02-91ER-40609.
Also, I gratefully acknowledge fruitful discussions with M.\ Gorenstein,
J.\ Lenaghan, I.\ Mishustin, O.\ Scavenius, D.J.\ Schwarz,
E.\ Shuryak, H.\ Sorge, N.\ Xu, and with
D.H.\ Rischke, whom I also thank for the permission to use parts of his
RHLLE code.


\end{document}